# Model predictive control of voltage profiles in MV networks with distributed generation

M. Farina, A. Guagliardi, F. Mariani, C. Sandroni, R. Scattolini

*Abstract*— The Model Predictive Control (MPC) approach is used in this paper to control the voltage profiles in MV networks with distributed generation. The proposed algorithm lies at the intermediate level of a three-layer hierarchical structure. At the upper level a static Optimal Power Flow (OPF) manager computes the required voltage profiles to be transmitted to the MPC level, while at the lower level local Automatic Voltage Regulators (AVR), one for each Distributed Generator (DG), track the reactive power reference values computed by MPC. The control algorithm is based on an impulse response model of the system, easily obtained by means of a detailed simulator of the network, and allows to cope with constraints on the voltage profiles and/or on the reactive power flows along the network. If these constraints cannot be satisfied by acting on the available DGs, the algorithm acts on the On-Load Tap Changing (OLTC) transformer. A radial rural network with two feeders, eight DGs, and thirty-one loads is used as case study. The model of the network is implemented in DIgSILENT PowerFactory®, while the control algorithm runs in Matlab®. A number of simulation results is reported to witness the main characteristics and limitations of the proposed approach.

## I. Introduction

The ever increasing diffusion of renewable distributed generators (DG) raises new technological problems in the management and control of Medium Voltage (MV) and Low Voltage (LV) distribution networks. In fact, distributed generators can induce local voltage increase, with inversion of power flows and emergence of inverse currents along the feeders of radial networks, so that voltage control is becoming of paramount importance for the further development of the field. On the other hand, the improved information and communication capabilities of modern Smart Grids (SG) allow to develop innovative control schemes and procedures which have to be fully exploited to guarantee a sustainable growth of distributed generation.

In recent years the voltage control problem has motivated many research efforts and a number of solutions have been proposed based on coordinated or uncoordinated schemes, see e.g. [1]-[10]. In coordinated solutions, a remote controller maintains prescribed voltage profiles and reactive power flows, see e.g. [2], [3], usually acting on an On-Load Tap Changer (OLTC) transformer, see [6], or on additional Energy Storage Systems (ESS) used to reduce the OLTC operations, see [10]. In uncoordinated schemes, the voltage and reactive power control equipments spread over the network are allowed to locally regulate the terminal bus voltage by adjusting their reactive power output, see [2]. Both coordinated and uncoordinated schemes have their own advantages and drawbacks. Coordinated control structures are potentially less reliable and prone to communication losses, local faults and overall vulnerability but, on the contrary, they usually rely on the solution of a global Optimal Power Flow (OPF) problem, as such they can reduce losses and optimize the voltage and reactive power profiles. Concerning uncoordinated solutions, the may display serious technical drawbacks, see again [2], [8]. However, the flexibility provided for example by photovoltaic generators with inverters, together with their wider and wider diffusion, will inevitably lead to their massive participation in voltage control in the future. In any case, and to the best of the authors knowledge, all the solutions proposed so far, both for centralized and distributed control structures, rely on a stationarity assumption, i.e. the network is assumed to be in permanent periodic regime. This prevents one from using classical dynamic control techniques, and the transient performance of the controlled system subject to disturbances are neglected, such as the transient response in front of varying loads or power production of the DGs.

In this paper a new dynamic model-based approach for voltage control in MV networks is proposed. The overall control system is designed according to a hierarchical (cascade) thee-layer structure, as suggested among the others in [4]. At the upper level, a static OPF problem is solved and the voltage and reactive power references at the nodes of the grid are computed. Based on the OPF solution, at the intermediate level a centralized controller computes the reference values for the local power factors of the distributed generators. Finally, at the lower level, these reference power factors are transformed into reference values of reactive power and local Automatic Voltage Regulators (AVR) are designed, one for each DG participating in the control action. Since the upper and the lower levels are quite standard, in the paper focus is placed on the design of the centralized controller at the intermediate level. Specifically, this controller is designed with a Model Predictive Control (MPC) algorithm based on an impulse response model of the network, which can be obtained with simple and non-invasive experiments on the real system or on a reliable and validated simulator. The choice of MPC is due to its capability to explicitly handle constraints on the main process variables, such as the voltages along the grid or the adopted power factors. In addition, by including suitable slack variables in the optimization problem underlying the MPC formulation, possible infeasibility conditions can be

M. Farina and R. Scattolini are with Dipartimento DEIB, Politecnico di Milano, Milano, Italy (e-mail: {marcello.farina;riccardo.scattolini}@polimi.it). F. Mariani is Ms student at the Politecnico di Milano (e-mail: marianifederico@gmail.com). C. Sandroni and A. Guagliardi are with RSE, Via Rubattino 54, Milano, Italy (e-mail: {guagliardi;sandroni}@rse-web.it)

detected, due for example to excessive load or generated power variations. In these cases, the tap changer position is modified to recover feasibility and to maintain the voltages inside the prescribed band.

The proposed approach has been used for control of a rural MV [20kV] radial network, located in the center of Italy, with two feeders, eight distributed generators and thirty-one loads. A detailed simulator of the network has been first developed in the DIgSILENT PowerFactory® environment and has been used to obtain the impulse response representation of the system. Then, the MPC algorithm has been implemented in the Matlab environment and has been tested on the DIgSILENT simulator.

The paper is organized as follows. In Section 2 the overall control structure is described and the MPC algorithm is introduced. Section 3 is devoted to present a number of simulation results. Finally, some conclusions and hints for future research are reported in Section 4.

## II. CONTROL STRUCTURE AND MPC DESIGN

The proposed control structure is made by three layers; at the upper layer (tertiary control) a static OPF is periodically solved and the optimal voltage profiles along the distribution network are computed based on the current status of the network and on the prediction of the future loads and active power production. At the intermediate layer (secondary control), a centralized MPC regulator computes the reference power factors for the DGs participating in voltage control, based on the solution of the OPF and on the available voltage measurements. At the lower layer (tertiary control), the reference power factors are transformed in reactive power references and local AVRs are used to control the reactive powers by acting on the excitation voltages of the DGs. A schematic representation of the control structure is shown in Figure 1.

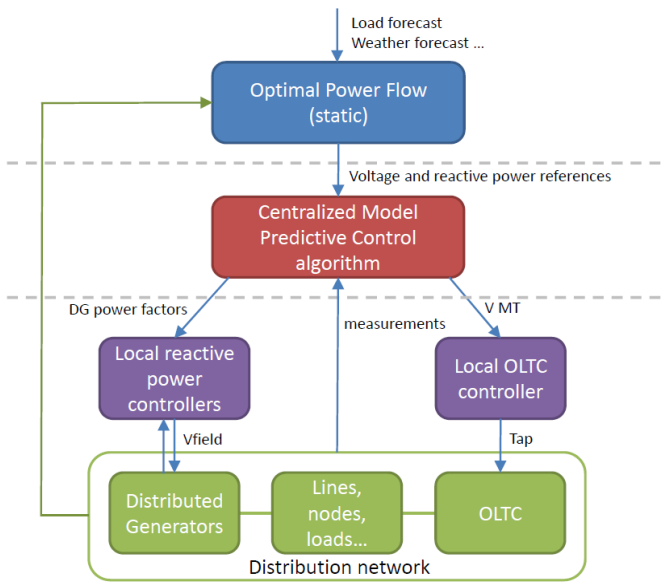

**Figure 1: The hierarchical control structure.**

In the following, focus is placed on the design of the centralized controller at the intermediate layer. In fact, well established techniques and tools are already available for OPF, see e.g. [12]. Moreover, the synthesis of the local AVR regulators, usually PI-PID, is in general not critical due to the satisfactory frequency decoupling of these control loops, see e.g. [13] where some preliminary results of this research have been reported.

### A. The MPC algorithm

The centralized controller at the intermediate layer is designed with MPC, a very popular method in the process industry which relies on the recursive solution of a constrained optimization problem, see [10]. The MPC approach has been selected for the following reasons:

• In the design phase it is possible to use impulse response models of the network, including the AVR control loops, which can be obtained by means of simple experiments on the real system or on a detailed dynamic simulator, such as the one developed with DigSILENT, a powerful and widely used industrial simulation environment. The main advantage of relying on input-output models, like the impulse response models used here, is that state estimators are not required. This is fundamental, in this framework, due to the presence of a large number of unknown and time-varying disturbances (usually larger than the number of measured variables), e.g., loads or produced power.

• Hard constraints can be easily included in the optimization problem to be solved on-line at any sampling time for the computation of the main control variables, i.e., the reference power factors of the DGs. This is a fundamental point, since in practical applications the zero-error asymptotic tracking of the voltage reference values at prescribed points of the grid is not the main issue. On the contrary, it must be guaranteed that these voltages remain inside the statutory limits, that reverse power flows are avoided, and that the operational constraints on the adopted power factors are fulfilled.

• Voltage deviations at specified nodes of the network can be differently weighted in the performance index to be recursively minimized, so that flexibility is easily achieved in the control problem formulation.

• Future predicted variations of the loads and of the power produced by some DGs (such as PVs) can be accounted for to enhance the control performance.

The adopted MPC algorithm is based on the following truncated linear discrete-time impulse representation of the system

$$y(k) = \sum_{i=1}^{M} [g_i u(k-i) + \gamma_i d(k-i)] + \delta(k)$$

where $k$ is the discrete time index, $y$ is the vector of the controlled variables (deviations of the voltages with respect to their nominal values at specified nodes of the grid), $u$ is the vector of the control variables (deviations of the power factor references), $d$ is the vector of measurable disturbances (possible known deviations of the active and the reactive power of the loads), $\delta$ is an additional unknown term summarizing the contribution of all the unknown exogenous signals (deviations of the unmeasurable loads) acting on the system, and $g_i$, $\gamma_i$ are the impulse response coefficients. It is assumed that the impulse response is practically exhausted after $M$ sampling times. The value of $\delta$ can be estimated as

$$\hat{\delta}(k) = y(k) - \sum_{i=1}^{M} g_i u(k-i) + \gamma_i d(k-i)$$

Accordingly, and assuming that $\delta$ is constant in the future, the *i*-th step ahead prediction $y(k+i)$ computed at time $k$ can be given the form

$$y(k+i)$$
$$= \sum_{j=1}^{i} (g_j u(k+i-j) + \gamma_j d(k+i-j)) + y(k)$$
$$+ \sum_{j=i+1}^{M} (g_j u(k+i-j) + \gamma_j d(k+i-j))$$
$$- \sum_{j=1}^{M} (g_i u(k-j) + \gamma_j d(k-j))$$

This prediction is a function of the past, present and future control actions and of the current output and it is used in the constrained optimization problem stated below. Defining

$$Y = \begin{bmatrix} y(k+1) \\ y(k+2) \\ \vdots \\ y(k+N) \end{bmatrix}, \quad U = \begin{bmatrix} u(k) \\ u(k+1) \\ \vdots \\ u(k+N_u-1) \end{bmatrix}$$

the MPC optimization problem is

$$\min_{U, \varepsilon_1, \varepsilon_2} Y'QY + U'RU + \mu_1 \varepsilon_1^2 + \mu_2 \varepsilon_2^2$$

subject to the following constraints on the input and output variables

$$U_{\min} \leq U \leq U_{\max}$$
$$\varepsilon_1 \mathbf{1} + Y_{\min} \leq Y \leq \varepsilon_2 \mathbf{1} + Y_{\max}$$
$$\varepsilon_1 \geq 0, \varepsilon_2 \geq 0$$

The matrices $Q$ and $R$ are positive definite and symmetric, $\mu_1 > 0$, $\mu_2 > 0$, and the prediction horizon $N$ must be selected to include the main system's dynamics. The parameter $N_u$ corresponds to the so-called control horizon, and is used in MPC to allow for only a limited number of variations of the future control variables, i.e. in the optimiziation problem it is set $u(k+N_u+i) = u(k+N_u-1)$, $i>0$. The terms $U_{\min}$, $U_{\max}$, $Y_{\min}$, $Y_{\max}$ represent physical bounds on the control and controlled variables, $\mathbf{1}$ is a vector of elements equal to one, and $\varepsilon_1$, $\varepsilon_2$ are so-called *slack variables*, introduced to allow soft constraints on the outputs, so as to guarantee feasibility also when disturbances (loads or generators variations) suddenly move the network away from its nominal operating conditions. This general formulation is very flexible, since additional constraints on the control variations at any sampling time, or on the maximum deviation allowed between two output (voltages) at adjacent nodes of the network, can be easily included to adapt the optimization problem to any specific requirement in terms of performance and constraints. As already noted, in the voltage control problem, these constraints have the highest importance, while the requirement of exact tracking of voltage reference values is less relevant. For this reason, no integral action has been included in the regulator structure.

The above optimization problem is *QP* (Quadratic Program), and can be efficiently solved at any time instant to compute the optimal sequence $u(k), \ldots, u(k+N_u-1)$ of future control variables. Then, according to the so-called Receding Horizon principle, only the first value $u(k)$ of this sequence is effectively applied and the overall procedure is repeated at the next sampling time.

### B. The OLTC controller

Large variations of the active power produced by the DG's and/or of the network loads can lead to the impossibility to maintain the voltages along the grid within the nominal operation bound $Y_{min} \leq Y \leq Y_{max}$ by solely modifying the DG's power factors using the MPC algorithm previously described. For a prompt response to this undesired situation, our regulation scheme is designed in such a way that the OLTC is activated when violations are revealed. Bound violations, in our framework, are detected when the slack variables included into the optimization problem to guarantee feasibility (i.e., $\varepsilon_1$, $\varepsilon_2$) take nonzero values. For this reason, the OLTC control has been implemented starting from the values taken by the slack variables. Specifically, letting $\delta\varepsilon = \varepsilon_2 - \varepsilon_1$, three cases can occurr and the proper control actions can be taken:

- $\delta\varepsilon = 0$ if the upper and lower constraints on the voltages along the feeders are satisfied or if they are violated with equal values $\varepsilon_1$ and $\varepsilon_2$. In both these cases, no variations of the position of the OLTC are forced.
- $\delta\varepsilon > 0$, if only the upper voltages constraints are violated ($\varepsilon_2 > 0$) or $\varepsilon_2 > \varepsilon_1$ . In this cases, the position of the tap selector is modified to reduce the voltage at the busbar (node N02 in the following     Figure **2**), and along the feeders.
- $\delta\varepsilon < 0$, if only the lower voltages constraints are violated ($\varepsilon_1 > 0$) or $\varepsilon_1 > \varepsilon_2$ . In this case, the position of the tap selector is modified to increase the voltage at the busbar (node N02 in the following    Figure **2**).

In any case, and in order to avoid high frequency switching, the previous conditions must be maintained for a prescribed "dwell time" $T_s$ before the tap position is allowed to switch for the first time or after a previous commutation.

### III. CASE STUDY

The MPC algorithm and the logic governing the tap position have been used to control the simulated model of the rural radial MV (20kV) network depicted in     Figure **2** and made by two feeders, eight DGs (photovoltaic, aeolic, turbogas), and thirty-one loads (industrial, agricultural, residential, tertiary, public lighting). For simplicity, all the DGs have been modeled as fifth order synchronous generators in the Park transformation domain. Feeder 1 is 27 km long with seventeen nodes and five DGs, while Feeder 2 is 36.9 km long with fourteen nodes and three DGs. A number of working points, corresponding to the load and generation profiles at hours 1a.m., 7a.m., 1p.m., 7p.m. of the day have been considered. The main characteristics of the DGs and of the loads in these working points, as well as of the network elements are reported in the Appendix.

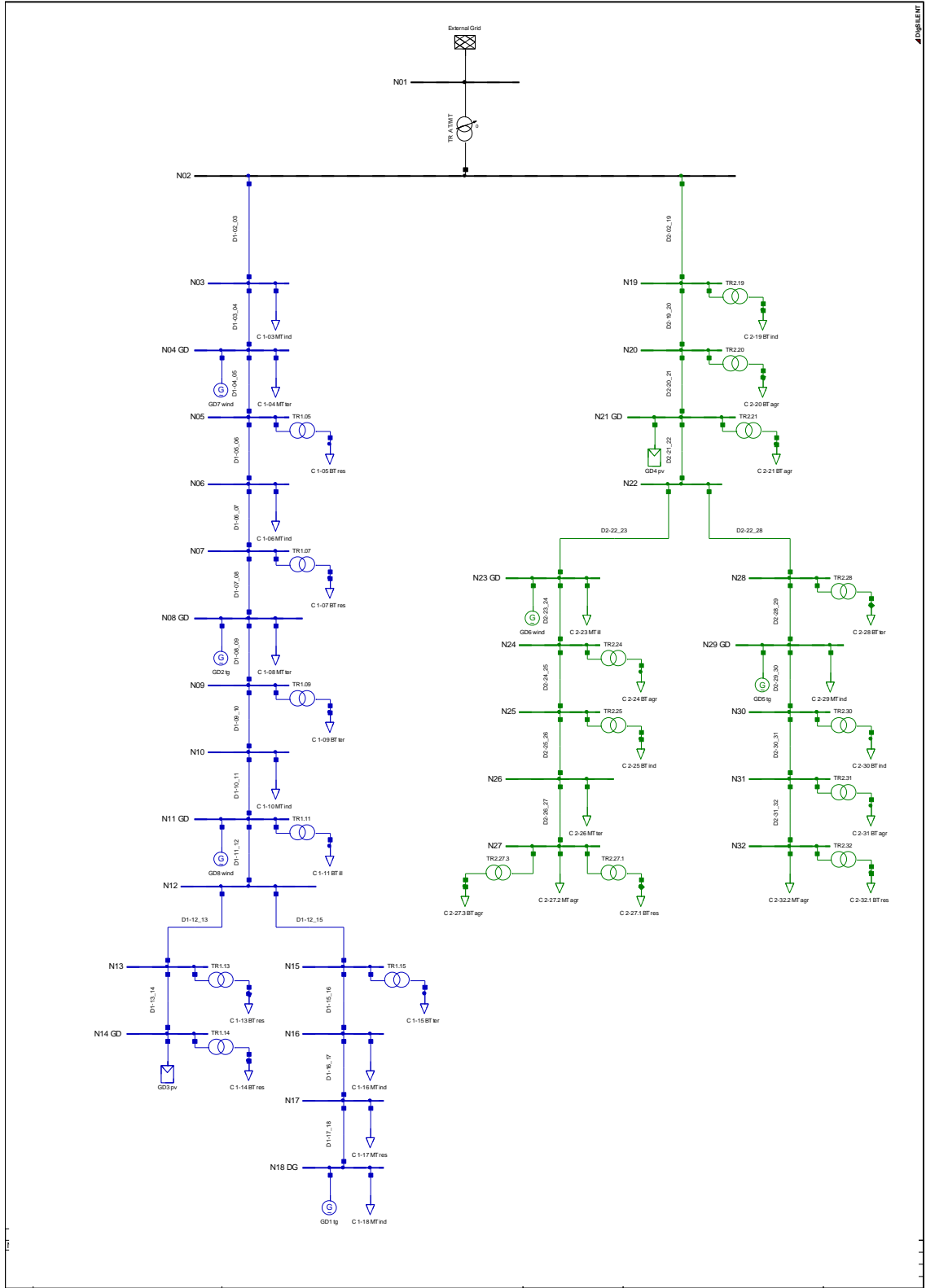

**Figure 2: the benchmark network.**

Note that the considered operating conditions are significantly different from each other in terms of power production of the DGs and of the loads.

A detailed simulator of the network has been first developed in the DIgSILENT environment. Then, by adopting the sampling time $T=2$s, this simulator has been used to obtain a discrete-time linearized model of the network at 7a.m. and based on an impulse response representation with $M=90$ regressors. Finally, simple PI-type AVR regulators have been tuned with empirical rules, one for each DG.

The MPC algorithm has been implemented in Matlab and an *ad hoc* software interface has been developed to link and synchronize the Matlab environment with the DigSILENT simulator, which plays the role of the controlled plant in the experiments described in the following. In the control algorithm, it has been assumed that that the controlled variables are eleven voltages, five corresponding to nodes N03, N06, N11, N14, N18 (see Figure **2**) of feeder 1 and six corresponding to nodes N19, N21, N23, N27, N28, N32 of feeder 2. The active and reactive powers of the distributed generators DG1, DG2, and DG3 have been considered as known disturbances, while the powers produced by the other DGs and all the loads have been assumed to be unknown disturbances. The prediction horizon $N=10$ has been chosen, while only two variations of the future control variables have been allowed, i.e. $N_u=2$. The weighting matrices have been set as $Q=10I$, $R=0.1$, where $I$ is the identity matrix of appropriate dimensions. Moreover, it has been set $\mu_1=\mu_2=1000$. The controlled voltages (in per units) have been constrained to belong to the interval [0.9 p.u, 1.1 p.u.], while the input variables, i.e. the reference power factors, have been constrained to belong to the interval [0.6,1].

*Experiment 1*

The performances of the MPC regulator, based on the impulse response model computed at 7a.m., have been tested in the four operating conditions specified in Table 2 and Table 8 in Appendix. Specifically, the perturbations listed in Table 1 have been given to some DGs and loads; note that their size is significantly large compared to the corresponding nominal values.

| Node | time | Variation |
|---|---|---|
| N32-2 | 20s | 50% increase of the load active and reactive power with respect to nominal values |
| DG2 | 100s | Active power step variation with final value 1.75 [MW] |
| N08 | 150s | 100% increase of the load active and reactive power with respect to nominal values |
| DG5 | 200s | Active power step variation with final value 3.75 [MW] |
| N16 | 300s | 50% reduction of the load active and reactive power with respect to nominal values |
| DG8 | 600s | Active power step variation with final value 3. [MW] |

**Table 1: Power variations of DGs and loads in Experiment 1.**

First, the network has been considered to be in the nominal stationary operating conditions at 7a.m., which however does not correspond to the desired equilibrium due to a too high voltage at node N18. Therefore, the regulator has the twofold objective to reach the required nominal operating point and to counteract the variations of Table 1. The transients of the voltages at the controlled nodes are reported in Figure 3, and show the excellent behavior of the controlled system: the upper and lower voltage constraints are satisfied (save for an initial transient due to the initial operating conditions outside these boundaries), and the voltage tend to reach the corresponding reference values. The control variables computed by MPC, i.e. the reference power factors of the DGs, are shown in Figure 4.

The ability of the MPC regulator, based on the model at 7a.m., to control the DigSILENT simulator of the network at different operating conditions has been tested starting from the nominal conditions at 1a.m, 1p.m. and 7p.m. and applying the same power variations summarized in Table 1. The obtained voltage profiles are reported in the following Figure 5 - Figure 10. These results show a slight deterioration of the performances due to the presence of an oscillatory behavior in many transients. However, the size of these oscillations is small, and could easily be smoothed with a simple filtering action on the implemented power factors. This has not been done here to fairly evaluate the performances of the control algorithm at operating conditions very different from those at 7a.m., where the network impulse response model has been derived. Note however that the voltage constraints are always met, save for the transient of N18 at 1a.m. (see Figure 9, upper panel), when no feasible solution exists for the stated optimization problem and the corresponding slack variable takes values different from zero. In order to improve these performances, it could be possible to adapt the MPC algorithm to varying operating conditions, as discussed in the final section of the paper.

*Experiment 2*

Considering again the conditions at 7 a.m., a simulation has been performed by disconnecting at time $t=180s$ the loads at nodes N03 and N18, both belonging to the first feeder. The transients of the controlled voltages with and without the OLTC controller, are compared in Figure 11 and Figure 12. In the implementation of the OLTC controller, a dwell time $T_s=75s$ has been used. It is apparent that the critical voltage at node N18 (see Figure 11) returns within the prescribed limits much faster when the OLTC action is available.

IV. CONCLUSIONS

The MPC algorithm proposed in this paper has been proved to be effective to control a complex benchmark describing a MV rural network working at the different operating conditions. This network represents a challenging test case, which calls for the solution of a large scale and multivariable control problem, with more controlled variables than manipulated inputs, and with more disturbances than manipulated variables. In addition, a physical model of the system would result to be too complex for the control design and stringent constraints must be met to cope with realistic conditions.

The proposed mixed control strategy, combining both the use of the OLTC and the participation of the DGs, led satisfactory results which can be further improved in many ways, for example by adopting some kind of adaptation mechanism. This can be done by simply modifying the weighting matrices appearing in the cost function at any new operating point according to a gain scheduling procedure. In addition, one could also rely on different linearized (impulse response) models determined at any new operating condition by means of the more complex and reliable simulator developed in DigSilent.

Further extensions of the proposed control scheme will deal with distributed implementations of the controller, see [14] for a survey of available distributed MPC methods, or on the use of a larger number of slack variables, one for each voltage constraint, to refine the control logic governing the OLTC.

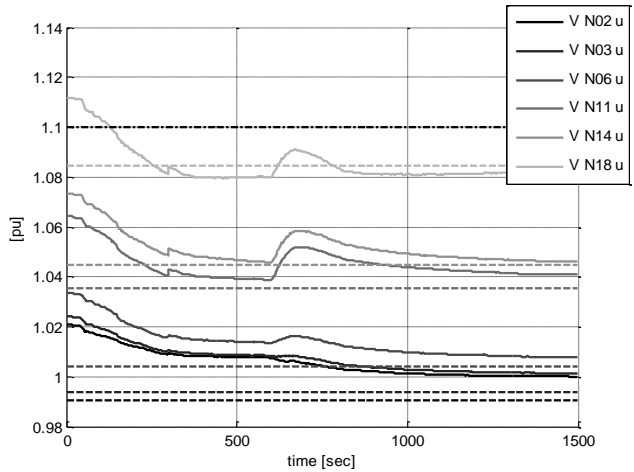
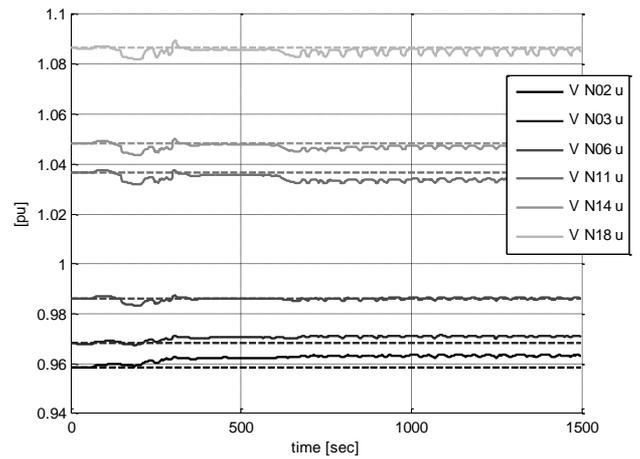

**Figure 3: Experiment 1 – 7a.m. voltages at the nodes of the first (up) and second (down) feeder. Dashed lines: reference values. Dash-dotted black line (upper panel): $Y_{max}$**

**Figure 5: Experiment 1 – 1p.m. voltages at the nodes of the first (up) and second (down) feeder. Dashed lines: reference values.**

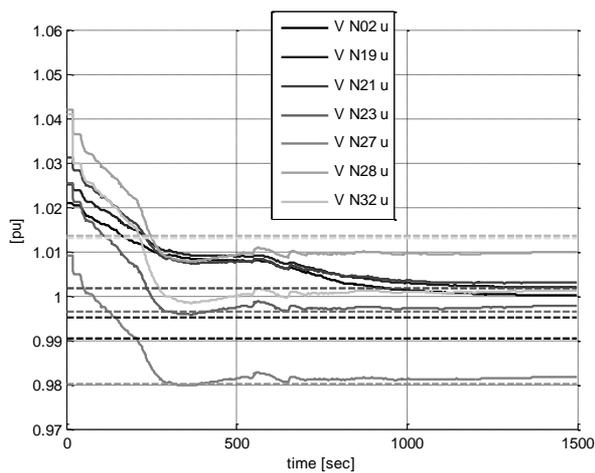
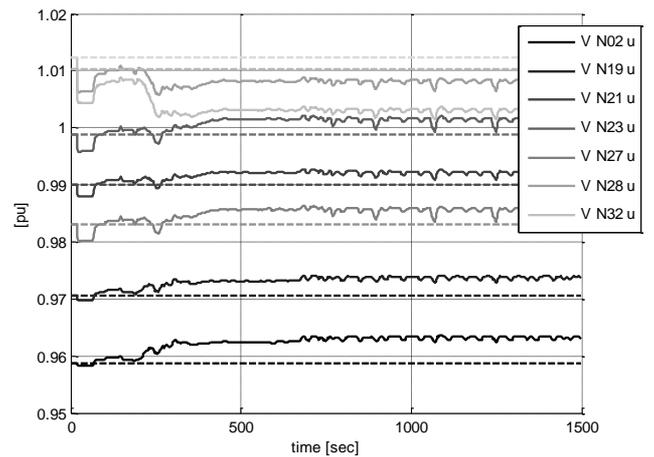

**Figure 4: Experiment 1 - 7a.m. reference power factors computed by the MPC algorithm.**

**Figure 6: Experiment 1 - 1p.m. reference power factors computed by the MPC algorithm.**

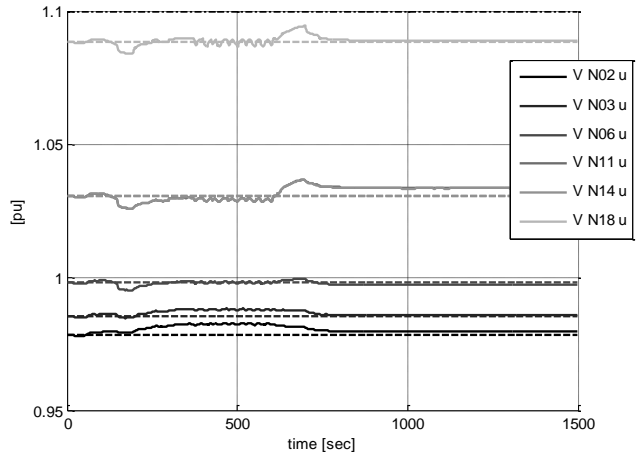
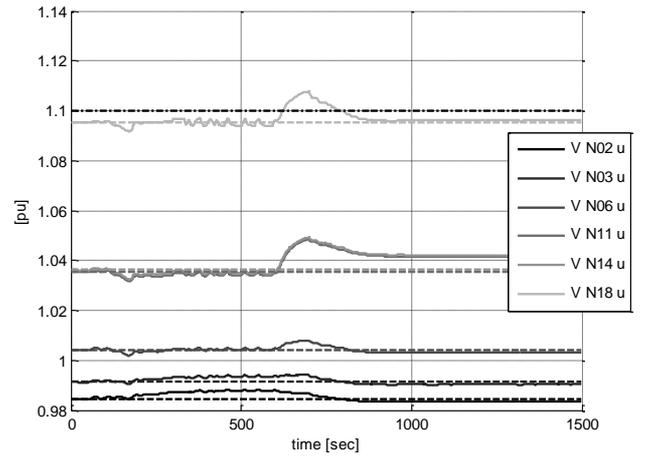
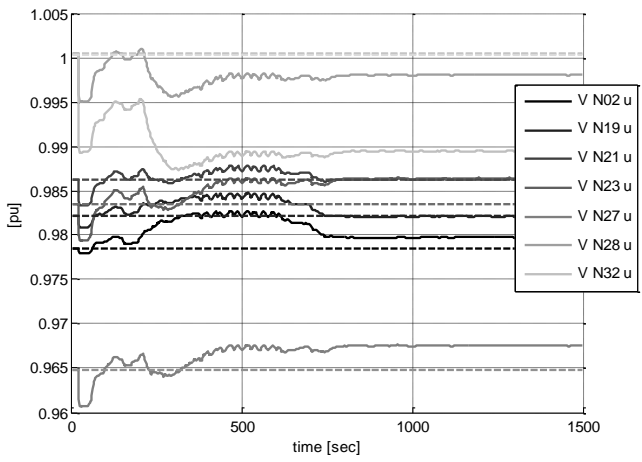
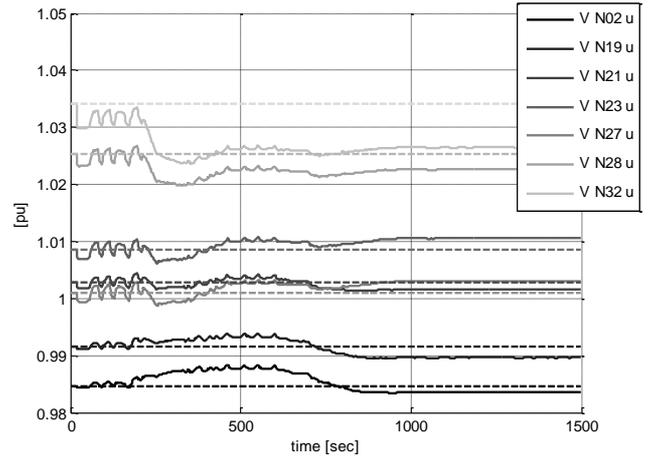

**Figure 7: Experiment 1 – 7p.m. voltages at the nodes of the first (up) and second (down) feeder. Dashed lines: reference values.**

**Figure 9: Experiment 1 – 1a.m. voltages at the nodes of the first (up) and second (down) feeder. Dashed lines: reference values. Dash-dotted black line (upper panel): $Y_{max}$**

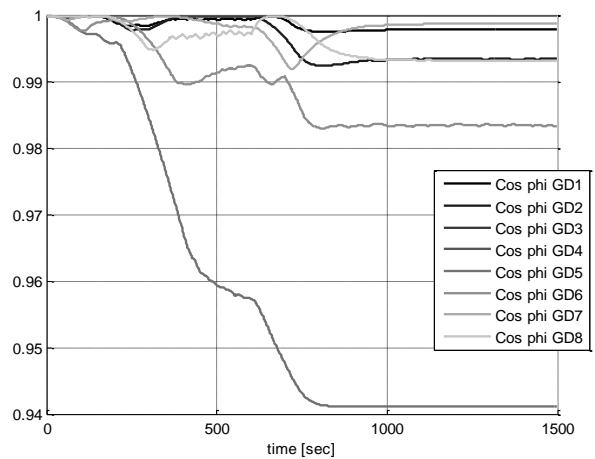
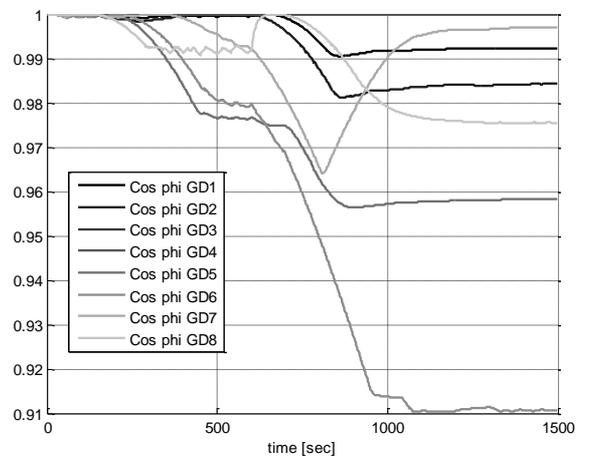

**Figure 8: Experiment 1 - 7p.m. reference power factors computed by the MPC algorithm.**

**Figure 10: Experiment 1 - 1a.m. reference power factors computed by the MPC algorithm.**

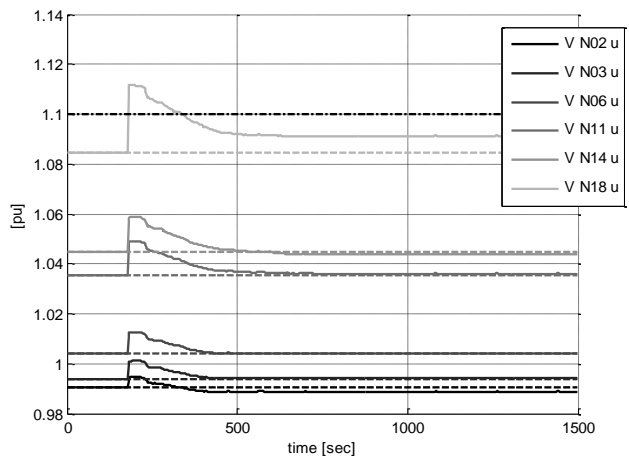

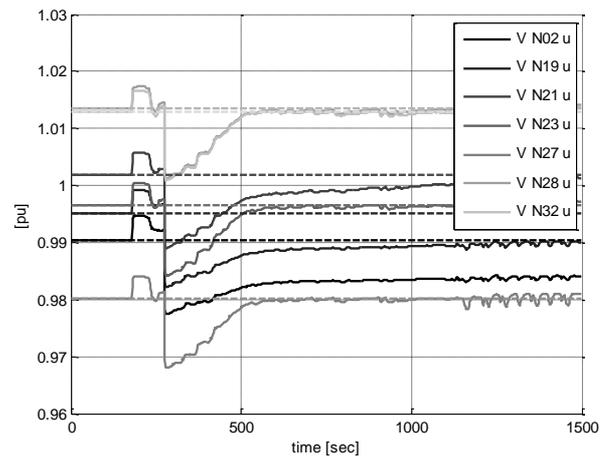

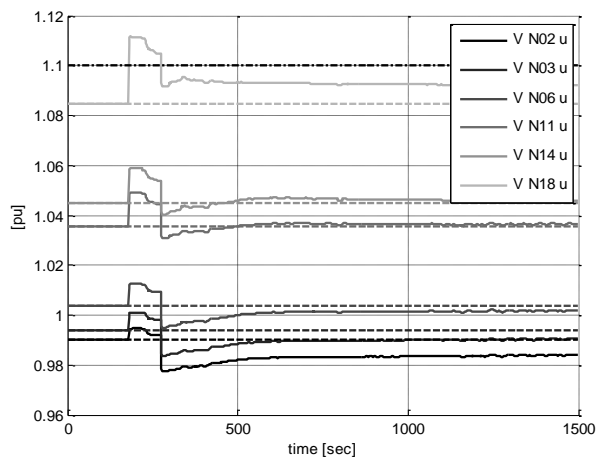

**Figure 11**: Experiment 4 - voltages at the nodes of the first feeder. Up: without OLTC control, down: with OLTC control. Dash-dotted black line: $Y_{max}$

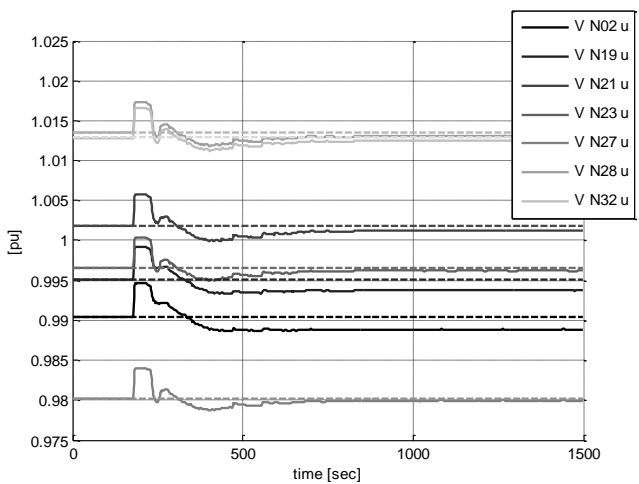

**Figure 12**: Experiment 4 - voltages at the nodes of the second feeder. Up: without OLTC control, down: with OLTC control.

APPENDIX

| DG | Feeder | P [MW] nominal | P [MW] 1a.m. | P [MW] 7a.m. | P [MW] 1p.m. | P [MW] 7p.m. |
|---|---|---|---|---|---|---|
| DG1 - TG | 1 | 5.5 | 4,95831 | 4,922 | 4,949614 | 4,964138 |
| DG2 - TG | 1 | 3.2 | 2,884827 | 2,864 | 2,879755 | 2,88806 |
| DG3 - PV | 1 | 3.2 | 0 | 1,559 | 2,056134 | 0 |
| DG4 - PV | 2 | 3.2 | 0 | 1,559 | 2,056124 | 0 |
| DG5 - TG | 2 | 5.5 | 4,958303 | 4,923 | 4,949595 | 4,963955 |
| DG6 - AE | 2 | 5.5 | 0,748335 | 0,549 | 3,245381 | 1,495977 |
| DG7 - AE | 1 | 5.5 | 0,823168 | 0,549 | 3,569903 | 1,645546 |
| DG8 - AE | 1 | 5.5 | 0,823169 | 0,549 | 3,569924 | 1,645603 |

**Table 2: distributed generators (PV: photovoltaic, TG: turbogas, AE: Aeolic).**

| Model | 40 MVA132/20 |
|---|---|
| Rated power | 50 MVA |
| Copper Losses | 176 kW |
| Relative Short-Circuit Voltage | 15.5 % |
| Number of taps | 12 (+6 … -6) |
| Voltage per tap | 1.5 % |

**Table 3: HV/MV transformer.**

| Model | 0.25 MVA 20kV/0.4 | 0.4 MVA 20kV/0.4 | 0.63 MVA 20kV/0.4 |
|---|---|---|---|
| Rated power | 250 kVA | 400 kVA | 630 kVA |
| Copper Losses | 2.6 kW | 3.7 kW | 5.6 kW |
| Relative Short-Circuit Voltage | 4 % | 4 % | 6 % |
| Number of transformers | 6 | 7 | 5 |

**Table 4: MV/LV transformers.**

| Name | Type |
|---|---|
| TR AT/MT | 40 MVA132/20 |
| TR1.05 | 0.63 MVA 20kV/0.4 |
| TR1.07 | 0.4 MVA 20kV/0.4 |
| TR1.09 | 0.25 MVA 20kV/0.4 |
| TR1.11 | 0.25 MVA 20kV/0.4 |
| TR1.13 | 0.25 MVA 20kV/0.4 |
| TR1.14 | 0.25 MVA 20kV/0.4 |
| TR1.15 | 0.25 MVA 20kV/0.4 |
| TR2.19 | 0.63 MVA 20kV/0.4 |
| TR2.20 | 0.4 MVA 20kV/0.4 |
| TR2.21 | 0.4 MVA 20kV/0.4 |
| TR2.24 | 0.4 MVA 20kV/0.4 |
| TR2.25 | 0.4 MVA 20kV/0.4 |
| TR2.27.1 | 0.63 MVA 20kV/0.4 |
| TR2.27.3 | 0.63 MVA 20kV/0.4 |
| TR2.28 | 0.25 MVA 20kV/0.4 |
| TR2.30 | 0.63 MVA 20kV/0.4 |
| TR2.31 | 0.4 MVA 20kV/0.4 |
| TR2.32 | 0.4 MVA 20kV/0.4 |

**Table 5: transformers.**

| Name | Type | Section [mm$^2$] | R [Ω/km] | L [mH/km] | C [uF/km] |
|---|---|---|---|---|---|
| ARG7H1RX 120mmq | Cable | 120 | 0,3330 | 0,382 | 0,2500 |
| ARG7H1RX 185mmq | Cable | 185 | 0,2180 | 0,350 | 0,2900 |
| ARG7H1RX 70mmq | Cable | 70 | 0,5800 | 0,414 | 0,2100 |
| Aerea Cu 25mmq | Overhead | 25 | 0,7200 | 1,389 | 0,0083 |
| Aerea Cu 70mmq | Overhead | 70 | 0,2681 | 1,286 | 0,0090 |

**Table 6: lines.**

| Name | Type | Length [km] | Feeder |
|---|---|---|---|
| D1-02_03 | ARG7H1RX 185 mmq | 1,884 | 1 |
| D1-03_04 | ARG7H1RX 185 mmq | 1,62 | 1 |
| D1-04_05 | ARG7H1RX 185 mmq | 0,532 | 1 |
| D1-05_06 | ARG7H1RX 185 mmq | 1,284 | 1 |
| D1-06_07 | ARG7H1RX 120 mmq | 1,618 | 1 |
| D1-07_08 | ARG7H1RX 120 mmq | 0,532 | 1 |
| D1-08_09 | ARG7H1RX 185 mmq | 2 | 1 |
| D1-09_10 | ARG7H1RX 185 mmq | 2,4 | 1 |
| D1-10_11 | ARG7H1RX 120 mmq | 2,252 | 1 |
| D1-11_12 | ARG7H1RX 185 mmq | 0,756 | 1 |
| D1-12_13 | Aerea Cu 25 mmq | 1,87 | 1 |
| D1-12_15 | ARG7H1RX 120 mmq | 1,19 | 1 |
| D1-13_14 | Aerea Cu 25 mmq | 1,28 | 1 |
| D1-15_16 | ARG7H1RX 120 mmq | 0,8 | 1 |
| D1-16_17 | Aerea Cu 25 mmq | 3 | 1 |
| D1-17_18 | Aerea Cu 25 mmq | 4 | 1 |
| D2-02_19 | ARG7H1RX 185 mmq | 3,6 | 2 |
| D2-19_20 | ARG7H1RX 185 mmq | 3,304 | 2 |
| D2-20_21 | Aerea Cu 70 mmq | 2,4 | 2 |
| D2-21_22 | Aerea Cu 70 mmq | 3,6 | 2 |
| D2-22_23 | Aerea Cu 70 mmq | 3 | 2 |
| D2-22_28 | ARG7H1RX 70 mmq | 2,4 | 2 |
| D2-23_24 | Aerea Cu 70 mmq | 3,08 | 2 |
| D2-24_25 | Aerea Cu 70 mmq | 1,65 | 2 |
| D2-25_26 | Aerea Cu 70 mmq | 1,8 | 2 |
| D2-26_27 | Aerea Cu 70 mmq | 2,2 | 2 |
| D2-28_29 | ARG7H1RX 70 mmq | 2,2 | 2 |
| D2-29_30 | ARG7H1RX 70 mmq | 2,4 | 2 |
| D2-30_31 | ARG7H1RX 70 mmq | 2,6 | 2 |
| D2-31_32 | ARG7H1RX 70 mmq | 2,7 | 2 |

**Table 7: lines characteristics.**

| LOAD | Type | P [MW] 1a.m. | Q [Mvar] 1a.m. | P [MW] 7a.m. | Q [Mvar] 7a.m. | P [MW] 1p.m. | Q [Mvar] 1p.m. | P [MW] 7p.m. | Q [Mvar] 7p.m. |
|---|---|---|---|---|---|---|---|---|---|
| N03 | I-MV | 0,4241 | 0,2056382 | 1,7007 | 0,8274 | 1,5251 | 0,7443214 | 0,512171 | 0,2506077 |
| N04 | T-MV | 0,0880 | 0,0435790 | 0,2206 | 0,1098 | 0,3982 | 0,2027317 | 0,398178 | 0,2021694 |
| N05 | R-LV | 0,1266 | 0,0849696 | 0,0919 | 0,0619 | 0,1865 | 0,1258367 | 0,223849 | 0,1504288 |
| N06 | I-MV | 0,2987 | 0,1466945 | 1,1963 | 0,5878 | 1,0766 | 0,534423 | 0,360688 | 0,178627 |
| N07 | R-LV | 0,0742 | 0,0505676 | 0,0537 | 0,0367 | 0,1101 | 0,0758206 | 0,131505 | 0,0899172 |
| N08 | T-MV | 0,0806 | 0,0421746 | 0,2018 | 0,1056 | 0,3667 | 0,20039 | 0,364454 | 0,1949992 |
| N09 | T-LV | 0,0224 | 0,0163072 | 0,0558 | 0,0401 | 0,1011 | 0,0749315 | 0,100247 | 0,07218664 |
| N10 | I-MV | 0,0690 | 0,0346640 | 0,2764 | 0,1388 | 0,2502 | 0,1283052 | 0,083363 | 0,04221028 |
| N11 | L-LV | 0,0848 | 0,0556931 | 0 | 0 | 0 | 0 | 0,069140 | 0,04507657 |
| N13 | R-LV | 0,0669 | 0,0463782 | 0,0488 | 0,0341 | 0,1016 | 0,0722708 | 0,118290 | 0,0821189 |
| N14 | R-LV | 0,0608 | 0,0422226 | 0,0446 | 0,0313 | 0,0929 | 0,0664264 | 0,107698 | 0,07484372 |
| N15 | T-LV | 0,0183 | 0,0140455 | 0,0455 | 0,0346 | 0,0831 | 0,0671084 | 0,081832 | 0,06273277 |
| N16 | I-MV | 0,1196 | 0,0609686 | 0,4786 | 0,2438 | 0,4341 | 0,2269448 | 0,144368 | 0,07421394 |
| N17 | R-MV | 0,2124 | 0,1096805 | 0,1525 | 0,0785 | 0,3193 | 0,1679388 | 0,378381 | 0,1965451 |
| N18 | I-MV | 0,2115 | 0,1131246 | 0,8431 | 0,4468 | 0,7641 | 0,414908 | 0,255114 | 0,1372592 |
| N19 | I-LV | 0,0528 | 0,0354222 | 0,2109 | 0,1401 | 0,1894 | 0,1265088 | 0,063739 | 0,04289173 |
| N20 | A-LV | 0,0570 | 0,0383210 | 0,1533 | 0,1028 | 0,1147 | 0,0774413 | 0,153274 | 0,1027773 |
| N21 | A-LV | 0,0572 | 0,0385852 | 0,1536 | 0,1032 | 0,1155 | 0,0782289 | 0,153287 | 0,1027909 |
| N23 | L-MV | 0,2321 | 0,1110559 | 0 | 0 | 0 | 0 | 0,182994 | 0,08827465 |
| N24 | A-LV | 0,0543 | 0,0366918 | 0,1439 | 0,0961 | 0,1094 | 0,0741473 | 0,143659 | 0,0958174 |
| N25 | I-LV | 0,0404 | 0,0274218 | 0,1598 | 0,1054 | 0,1450 | 0,0984401 | 0,048166 | 0,03204506 |
| N26 | T-MV | 0,0883 | 0,0442469 | 0,2174 | 0,1033 | 0,3993 | 0,2049369 | 0,388592 | 0,1829258 |
| N27 | R-LV | 0,1685 | 0,1128906 | 0,1193 | 0,0787 | 0,2471 | 0,1656134 | 0,285200 | 0,1853766 |
| N27.2 | A-MV | 0,3021 | 0,1470421 | 0,8023 | 0,3866 | 0,6083 | 0,2971269 | 0,799597 | 0,384729 |
| N27.3 | A-LV | 0,1048 | 0,0704947 | 0,2744 | 0,1814 | 0,2094 | 0,1407233 | 0,273508 | 0,1805081 |
| N28 | T-LV | 0,0270 | 0,0197459 | 0,0664 | 0,0464 | 0,1205 | 0,08741 | 0,118186 | 0,08054955 |
| N29 | I-MV | 0,0434 | 0,0220095 | 0,1726 | 0,0864 | 0,1561 | 0,0798601 | 0,051826 | 0,02596448 |
| N30 | I-LV | 0,0541 | 0,0378280 | 0,2135 | 0,1453 | 0,1934 | 0,1347733 | 0,064483 | 0,04441056 |
| N31 | A-LV | 0,0467 | 0,0319818 | 0,1234 | 0,0835 | 0,0934 | 0,0640347 | 0,123344 | 0,08350521 |
| N32.1 | R-LV | 0,0910 | 0,0630751 | 0,0643 | 0,0439 | 0,1333 | 0,0923274 | 0,155504 | 0,1052517 |
| N32.2 | A-MV | 0,2807 | 0,1388702 | 0,7457 | 0,3652 | 0,5634 | 0,2791258 | 0,745457 | 0,3650585 |

**Table 8: loads – type: A=Agricultural, R= residential, T=tertiary, I= industrial, L=public lighting, LV=low voltage, MV=medium voltage**